\documentclass[preprint2]{aastex6}
\usepackage{color}
\usepackage[titletoc]{appendix}
\usepackage[fleqn]{amsmath}
\usepackage{float}
\usepackage{threeparttable}
\usepackage{comment}
\usepackage{enumitem}
\usepackage{natbib}

\shortauthors{Millholland et al.}
\shorttitle{Same Star -- Same Planets}

\begin{document} 

\title{\textit{Kepler} Multi-Planet Systems Exhibit Unexpected Intra-system \\ Uniformity in Mass and Radius} 
\author{Sarah Millholland$^{1, 2}$, Songhu Wang$^{1, 3}$, Gregory Laughlin$^{1}$}
\altaffiliation{$^2$ NSF Graduate Research Fellow}
\altaffiliation{$^{3}$ 51 Pegasi b Fellow}
\affil{$^1$ Department of Astronomy, Yale University, New Haven, CT 06511} 
\email{sarah.millholland@yale.edu}

\begin{abstract}
The widespread prevalence of close-in, nearly coplanar super-Earth- and sub-Neptune-sized planets in multiple-planet systems was one of the most surprising results from the \textit{Kepler} mission. By studying a uniform sample of \textit{Kepler} ``multis'' with mass measurements from transit timing variations (TTVs), we show that a given planetary system tends to harbor a characteristic type of planet. That is, planets in a system have both masses and radii that are far more similar than if the system were assembled randomly from planets in the population. This finding has two important ramifications. First, the large intrinsic compositional scatter in the planet mass-radius relation is dominated by system-to-system variance rather than intra-system variance. Second, if provided enough properties of the star and primordial protoplanetary disk, there may be a substantial degree of predictability in the outcome of the planet formation process. We show that stellar mass and metallicity account for of order $20\%$ of the variation in outcomes; the remainder is as-yet unknown. 

\end{abstract}

\section{Introduction}

NASA's \textit{Kepler} mission \citep{2010Sci...327..977B} discovered an abundance of close-in ($P \lesssim 100$ days), tightly-spaced multiple-planet systems \citep{2011ApJS..197....8L, 2011ApJ...732L..24L, 2014ApJ...784...44L, 2014ApJ...784...45R}. These systems are nearly coplanar \citep[e.g.][]{2012ApJ...761...92F}, have low eccentricities \citep{2015ApJ...808..126V}, and are rarely in low-order mean-motion resonances \citep[e.g.][]{2014ApJ...790..146F, 2015MNRAS.448.1956S}. Recent work by \cite{2017arXiv170606204W} has further revealed that these multi-transiting systems tend to be comprised of planets that are regularly-spaced and more similarly-sized than planets drawn randomly from the aggregate population. Simultaneous to staying similarly-sized, the radii also tend to increase slightly with orbital separation \citep{2017arXiv170606204W, 2018MNRAS.473..784K}. 

A key question is whether this statistically significant ``peas-in-a-pod'' trend in planet radii extends to masses as well. Most planets comprising the \textit{Kepler} multis are so-called super-Earth/sub-Neptune type planets with $R_p = 1-4 R_{\oplus}$. These planets are common, occurring around roughly half of Sun-like stars \citep{2013PNAS..11019273P, 2013ApJ...766...81F}. Because they bridge the gap between rocky and volatile-rich, super-Earths/sub-Neptunes exhibit a startlingly wide range of densities and plausible compositions \citep[e.g.][]{2010ApJ...712..974R}.  The variance in possible gas fractions yields planet radii that can be largely independent of mass \citep{2014ApJ...792....1L} and produces significant intrinsic scatter in the mass-radius relation \citep{2014ApJ...783L...6W, 2015ApJ...806..183W, 2016ApJ...825...19W}. 

Given \cite{2017arXiv170606204W}'s observation of radius uniformity in multiple-planet systems, there are two options for the masses, each of which is unusual and consequential. Either the masses are dissimilar and the planets in a system are somehow capable of coordinating their radii, or the masses -- like the radii -- are similar, so that for a given system, the scatter in the mass-radius relation is significantly reduced. Our goal here is to determine which one of these scenarios dominates.   

Most \textit{Kepler} multis consist of stars that are too dim and/or planets that are too small and distant for reliable radial velocity (RV) mass measurements. However, if the planets in a system exhibit significant inter-planetary gravitational perturbations -- most often by being near a mean-motion resonance -- the perturbations can add constructively and produce substantial transit-timing variations (TTVs)  \citep{2005MNRAS.359..567A, 2005Sci...307.1288H}. Measuring TTVs for a multiple-planet system permits estimation of the planet masses and eccentricities (see \citealt{2017arXiv170609849A} for a review).

Recently, \cite{2017AJ....154....5H} conducted a comprehensive analysis of  \textit{Kepler} multiple-planet systems exhibiting significant TTVs (55 systems in total) to derive mass estimates. This dataset offers a unique opportunity to assess the degree of intra-system similarity of planet masses within \textit{Kepler} multis.\footnote{Our full analysis is available at:\\ \href{https://github.com/smillholland/Kepler_Multis_Uniformity}{https://github.com/smillholland/Kepler\_Multis\_Uniformity}.} 

\begin{figure}
\vspace{0cm}\hspace{0cm}
\includegraphics[width=\columnwidth]{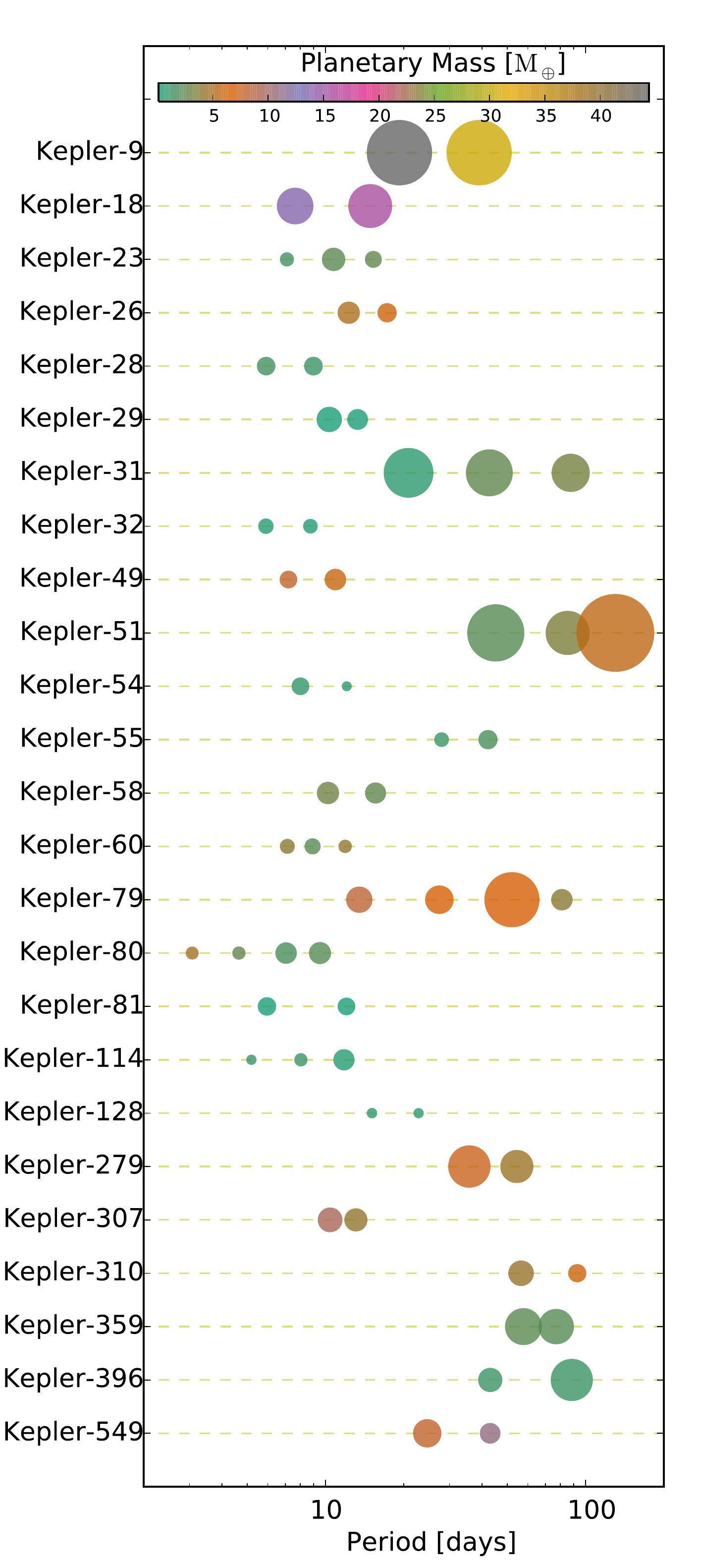}
\centering
\caption{Orbital architectures for 25 of the 37 systems in our sample. The dots are sized according to the planets' relative radii and colored according to mass. The planet with the smallest radius in this figure is Kepler-54 c with $1.3 \ R_{\oplus}$; the largest is Kepler-51 d with $10.1 \ R_{\oplus}$. Visual inspection suggests a high degree of within-system uniformity in both radii and masses.}
\label{Ten_best_systems}
\end{figure}

\section{System Clustering and Ordering in the Mass-Radius Plane}

We first examine the arrangement of \textit{Kepler} multiple-planet systems in the mass-radius plane. \cite{2017arXiv170606204W} already demonstrated that planets in the same system tend to be similar in radius. We ask: Does this hold for mass too? 

Our sample selection begins with the 145 \textit{Kepler} planets (55 systems) with masses derived by the TTV analysis conducted by \cite{2017AJ....154....5H}. They used two prior distributions for the masses: ``default'' and ``high mass''. We restricted the sample to the 37 systems for which all planets in a system had default and high mass estimates that agreed within $2\sigma$ uncertainties, and we adopted the default estimates. 
The 37 multi-planet systems contain 89 planets; there are 26 systems with two planets, 8 with three, 2 with four, and 1 with five. Planet radii and host star properties were taken from the California-Kepler survey (CKS) catalog\footnote{\href{https://california-planet-search.github.io/cks-website/}{https://california-planet-search.github.io/cks-website/}} \citep{2017AJ....154..107P, 2017AJ....154..108J}; the CKS parameters were available for 73 of the 89 planets. When they weren't available, or when the CKS and  \cite{2017AJ....154....5H} planet radii deviated by more than $100\%$ (as they did in 4 cases), we used the \cite{2017AJ....154....5H} radii for the discrepant planet and all others in the system. We also updated the planet mass estimates from \cite{2017AJ....154....5H} by using the CKS catalog stellar masses.

Figure~\ref{Ten_best_systems} displays the orbital architectures of 25 of the 37 systems, with points scaled to planet radii and coloration according to mass. The systems were selected based on a metric to be defined in the next section. Planets in a given system tend to be remarkably similar, not only in radius, but also in mass. In the following section, we will rigorously quantify this tendency for all 37 systems as an aggregate population.

\subsection{Clustering in the $M_p$--$R_p$ plane}
\label{clustering}

\begin{figure*}
\epsscale{1.2}
\plotone{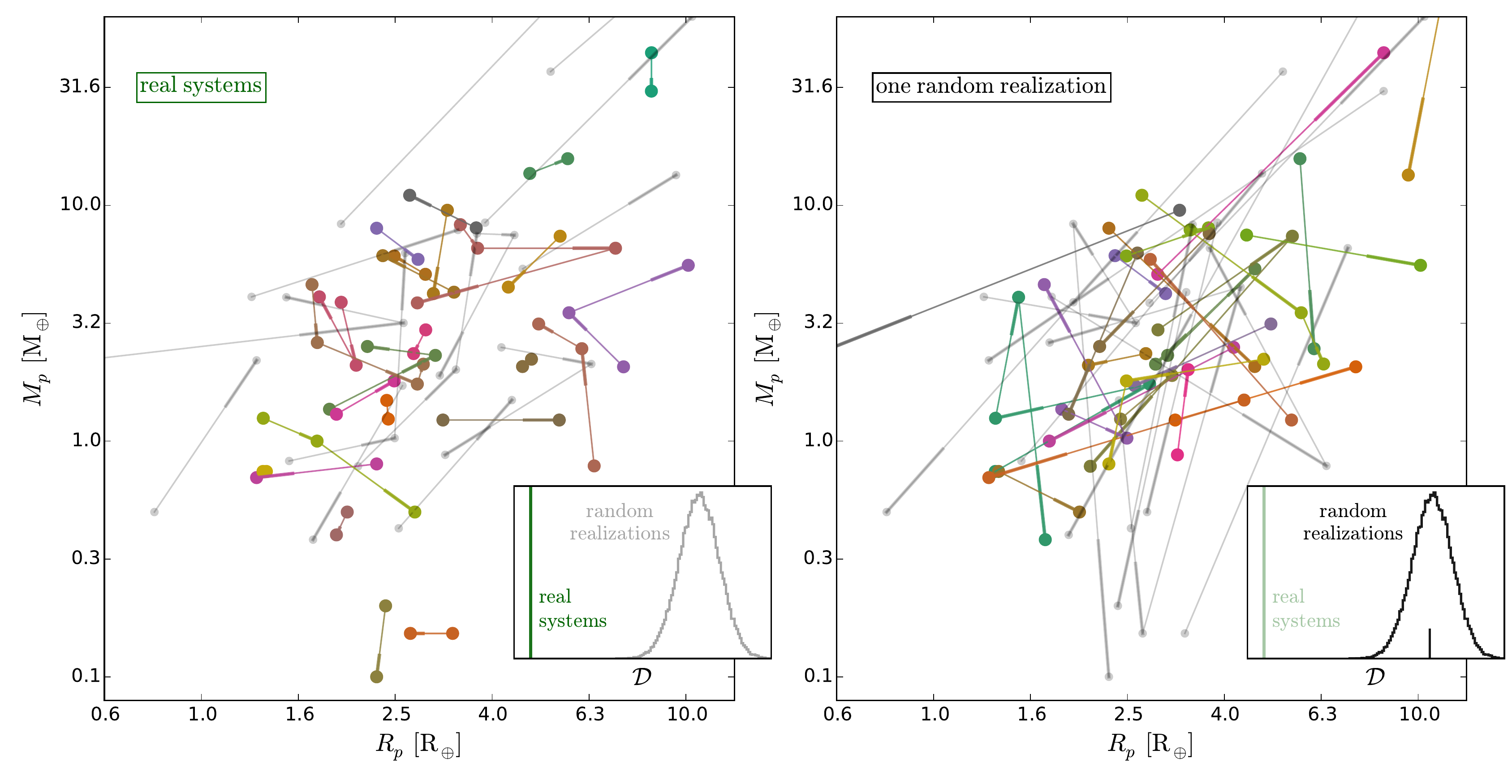}
\caption{The planet sample plotted as directed graphs in $M_p$--$R_p$ space. All planets in a given system have the same color and are connected by lines in orbital period order, with the thicker part of each segment attached to the planet with the larger period. For clarity, only the 25 systems with the smallest normalized directed graph distance are brightly colored. The left panel displays the 37 real systems, and the right panel depicts the directed graphs associated with a random permutation of the planets in which the number of systems and number of planets in each system are conserved, but where planet mass/radius pairs are shuffled. (See text.) The median radius uncertainty of the sample is $0.3 \ R_{\oplus}$, and the median mass uncertainty is $0.7 \ M_{\oplus}$ (low) and $1.5 \ M_{\oplus}$ (high). Insets in the lower right compare the dispersion metric (see equation~\ref{dispersion metric}) between the set of real systems (vertical green line) and the control population of 50,000 realizations of shuffled systems (histogram). The random realization depicted in the right panel represents just one member of the histogram; its location is marked with a black tick.} 
\label{Mp_vs_Rp_directed_graphs}
\end{figure*}

One way to display intra-system clustering in the $M_p$--$R_p$ plane is with a set of directed graphs.   The left panel of Figure~\ref{Mp_vs_Rp_directed_graphs} uses a directed graph for each of the 37 systems. All planets in a given system have the same color and have lines connecting them in orbital period order. For clarity, the brightly-colored systems are the 25 with the smallest normalized total distance of the graph, where the normalization is by the number of planets in the system. The remaining systems are plotted in light gray. Inspection of the left panel of Figure~\ref{Mp_vs_Rp_directed_graphs} makes it clear that planets in a given system tend to be closer together than expected if the systems were randomly assembled from planets in the population. That is, if we know the location of one planet on the $M_p$--$R_p$ plane, we can guess with better than random accuracy where the other planets in the system will fall.


Although this phenomenon of low $M_p$ and $R_p$ system dispersion is identifiable by eye, a quantitative metric is necessary for a rigorous analysis. The lengths of the lines in Figure~\ref{Mp_vs_Rp_directed_graphs} are a good measure of the dispersion of an individual system. To measure all systems at once, we can consider the total distance, $\mathcal{D}$, of all systems' directed graphs in $\log M_p$--$\log R_p$ space. This may be expressed as
\begin{equation}
\label{dispersion metric}
\mathcal{D} = \sum_{i=1}^{N_{\mathrm{sys}}} \sum_{\substack{j=1 \\ P_j < P_{j+1}}}^{N_{\mathrm{pl}}-1} \left[ \left(\log\frac{R_{j+1}}{R_j}\right)^2 + \left(\log\frac{M_{j+1}}{M_j}\right)^2 \right]^{1/2}.
\end{equation}
Calculating the distance in log-space is beneficial because the resulting metric involves intra-system ratios of radii and masses, which are known to better precision than absolute values. In part, this is due to the elimination of the stellar radius and mass uncertainties. An additional boost comes from the fact that TTV-derived planet mass ratios are often more tightly constrained than individual masses \citep[e.g.][]{2016ApJ...820...39J, 2017AJ....154....5H}.

This metric for system clustering on the $M_p$--$R_p$ plane is only meaningful if provided a control population for comparison. To this end, we constructed 50,000 control sets of shuffled systems. In each realization, the total number of systems and the number of planets in each system were conserved, but the planets themselves were randomly shuffled. To denote this, let $\mathcal{M} = (M_1, M_2, ..., M_{N})$ and $\mathcal{R} = (R_1, R_2, ..., R_{N})$ be the set of all planet masses and radii in the population. We define the symmetric group, $S_{N}$, as the set of all permutations of the set, $\{1, 2, ..., N\}$. If $X \in S_{N}$, then $\mathcal{M}(X)$ and $\mathcal{R}(X)$ represent a random shuffling of the planets, where mass/radius pairs are maintained. The set of distance metrics for the control population is therefore given by $\{\mathcal{D}(X) = \mathcal{D}[\mathcal{M}(X), \mathcal{R}(X)] \ \mid \ X \in S_{N}\}$.
The right panel of Figure~\ref{Mp_vs_Rp_directed_graphs} shows the directed graphs for one realization of randomly shuffled systems. Just like the left panel, the 25 systems with the smallest normalized total distances are plotted in color, with the others in gray. It is clear that the real systems are much less dispersed than the systems in this random realization. For instance, most of the gray systems in the left panel are significantly more clustered than the average among the colored, best 25 in the right panel. 

The panels in Figure~\ref{Mp_vs_Rp_directed_graphs} display the histogram of $\{\mathcal{D}(X) \mid \ X \in S_{N}\}$ for the 50,000 sets of shuffled systems. The value of $\mathcal{D}$ for the real systems is indicated with a vertical green line. The distance (or equivalently, dispersion) metric for the real systems is quite remarkably small in comparison to the distribution of $\{\mathcal{D}(X) \mid \ X \in S_{N}\}$ for the control population. It is smaller than the mean of the control set by $8.3\sigma$. 
We can also consider the individual mass and radius components of the distance, $\mathcal{D}_R$ and $\mathcal{D}_M$. Specifically we define 
\begin{equation}
\label{radius dispersion metric}
\mathcal{D}_R = \sum_{i=1}^{N_{\mathrm{sys}}} \sum_{\substack{j=1 \\ P_j < P_{j+1}}}^{N_{\mathrm{pl}}-1} \bigg| \log\frac{R_{j+1}}{R_j} \bigg|,
\end{equation}
with a similar expression for $\mathcal{D}_M$. The values for the real systems for radius and mass are respectively $5.3\sigma$ and $7.6\sigma$ smaller than the control population means.

The extremely statistically significant reduction in the system dispersion metric, $\mathcal{D}$, compared to the random expectation definitively indicates that planets in systems tend to be comparable to one another in mass and radius. Among various important implications, this is of prime relevance for understanding the arrangement of planets in the mass-radius diagram. 

Many authors have noted that the mass-radius relation exhibits astrophysical or intrinsic scatter \citep{2014ApJ...783L...6W, 2015ApJ...806..183W, 2016ApJ...825...19W}. That is, at a particular radius, there is a wide distribution of possible masses, reflecting a diversity in compositions. This scatter is particularly pronounced for super-Earths/sub-Neptunes \citep{2014ApJ...792....1L}. \textit{Our conclusion here is that astrophysical scatter in the mass-radius relation is dominated by -- or at least largely linked to -- system-to-system compositional dispersion rather than within-system dispersion.} 

\subsection{Ordering in $M_p$ and $R_p$}
\label{ordering}

Another property of interest is intra-system radius and mass ordering when the planets are ranked by orbital period. In a recent paper, \cite{2018MNRAS.473..784K} examined the radius size-ordering of observed \textit{Kepler} multiple-planet systems and found that systems tend to be statistically significantly size-ordered (thereby occupying low-entropy configurations) when compared to random expectation. A schematic way of investigating this for the dataset under consideration is to plot all of the $M_p$--$R_p$ directed graphs in Figure~\ref{Mp_vs_Rp_directed_graphs} end-to-end. Drawing an analogy to a random walk, the first planet in a given system is co-located with the last planet in the preceding system. 

\begin{figure}
\epsscale{1.2}
\plotone{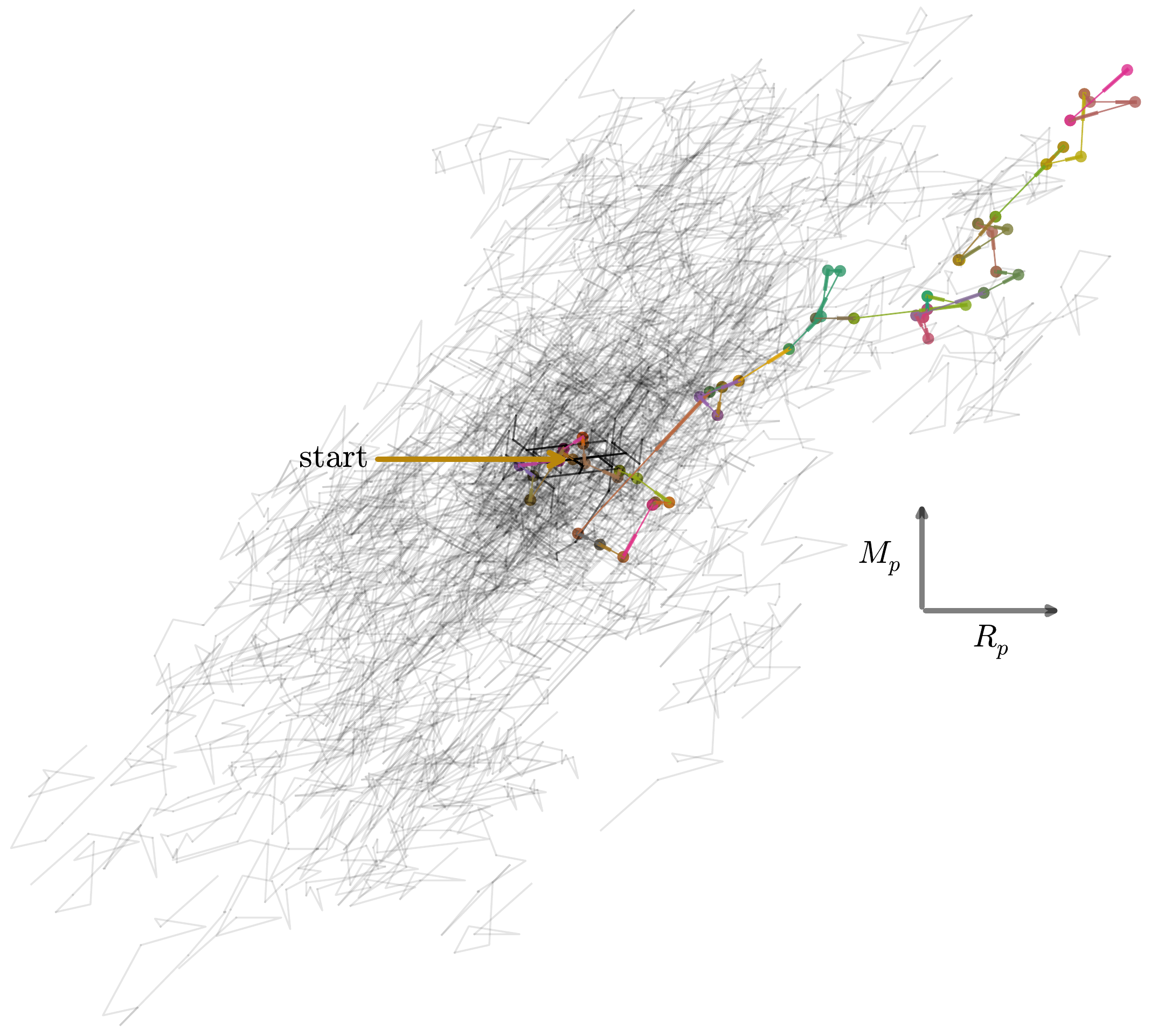}
\caption{``Random walks'' generated by directed graphs in the $M_p$--$R_p$ plane placed end-to-end. The colored trajectory corresponds to the real systems, sorted from low to high stellar mass. The black trajectories correspond to 100 sets of systems with a randomized order for the planets in each system. } 
\label{Mp_vs_Rp_directed_graphs_random_walk}
\end{figure}

Figure~\ref{Mp_vs_Rp_directed_graphs_random_walk} presents this schematic. The colored line represents the random walk of the set of 37 real systems, ordered from low to high stellar mass. The black lines represent the random walks of 100 sets of shuffled systems. These sets were randomized slightly differently than in Section~\ref{clustering} because here we are isolating intra-system mass and radius ordering. Rather than assembling the systems from permutations of all planets in the population, we simply randomized the order of the planets in each system for each realization. 

The trajectory for the real systems tends upwards and to the right, confirming that, on average, the planets in these systems tend to be size-ordered in both mass and radius. At its end, the trajectory is also at the edge of the ``cloud'' of the control population. We can quantify ordering metrics, $\mathcal{O}_R$ and $\mathcal{O}_M$, as the total displacement of the trajectories in the radius and mass directions, respectively. Specifically, we define $\mathcal{O}_R$ as
\begin{equation}
\label{radius ordering metric}
\mathcal{O}_R = \sum_{i=1}^{N_{\mathrm{sys}}} \sum_{\substack{j=1 \\ P_j < P_{j+1}}}^{N_{\mathrm{pl}}-1} \log\frac{R_{j+1}}{R_j}, 
\end{equation}
with a similar expression for $\mathcal{O}_M$. Considering 50,000 sets in the control population, we find that $\mathcal{O}_R$ for the real systems is $2.4\sigma$ larger than the control mean, and $\mathcal{O}_M$ is $1.8\sigma$ larger.\footnote{If we use all 55 \cite{2017AJ....154....5H} systems, the results for $\mathcal{D}$, $\mathcal{D}_R$, $\mathcal{D}_M$, $\mathcal{O}_R$ and $\mathcal{O}_M$ are, respectively, $9.6\sigma$, $8.2\sigma$, $8.2\sigma$, $3.0\sigma$ and $1.7\sigma$.}

Although this is not the first time that planet size ordering in the \textit{Kepler} multis has been shown \citep{2013ApJ...763...41C, 2017arXiv170606204W, 2018MNRAS.473..784K}, here we have not only confirmed the observation of radius ordering but also extended it to mass.





\section{Correlations with Host Star Properties}

We have shown that \textit{Kepler} multi-transiting systems tend to host ``the same'' planets, that is, planets that are self-similar in their radii and masses. Consequently, if provided all parameters of a host star and its natal protoplanetary disk, it should be possible (hypothetically speaking) to predict the most probable outcome of planet formation. What sets this default outcome? Here we use regression analyses to investigate which characteristics of host stars might be important in controlling the properties of their planetary systems.

We first expand our sample to a much larger collection of \textit{Kepler} multis rather than focusing on the 37 from \cite{2017AJ....154....5H}.  The trends in mass uniformity have already been examined, so we withhold planet masses from the analysis and focus instead on system characteristics involving orbital and planetary radii. The \cite{2017arXiv170606204W} ``CKS multis'' form our base dataset. For consistency, we applied the same cuts as \cite{2017arXiv170606204W} and discarded known false positives, systems in \cite{2017AJ....153...71F} with dilution from nearby stellar companions greater than $5\%$, grazing transits with impact parameters, $b > 0.9$, and planets with $\mathrm{SNR} < 10$. The total sample contains 909 planets in 355 multi-planet systems. Before proceeding, we report the values of the CKS multis intra-system radius dispersion and ordering metrics. We find that $\mathcal{D}_R$ and $\mathcal{O}_R$ are $14.9\sigma$ smaller/$10.5\sigma$ larger than their respective control population means.

Multiple regression is well-suited to the problem of correlating planet system characteristics with stellar properties. We used the the ordinary least squares, general linear regression from the \texttt{scikit-learn} Python machine learning package \citep{scikit-learn}\footnote{\href{http://scikit-learn.org/stable/modules/linear_model.html}{http://scikit-learn.org/stable/modules/linear\_model.html}}. The stellar properties (independent variables) that are potentially of interest and provided in the CKS catalog are $M_{\star}$, $R_{\star}$, $T_{\mathrm{eff}}$, metallicity, $\log g$, stellar age, and $v \sin i$. Due to their strong correlations with $M_{\star}$, we ignored $R_{\star}$, $\log g$, and $T_{\mathrm{eff}}$ to avoid complications from multicollinearity. 

For summary planetary system characteristics, we considered the semi-major axis and equilibrium temperature of the innermost planet (quantifying the ``inner edge'' of the system), the median planet radius, and a measure of dispersion of the planet radii, $\mathrm{CV}_{R_p}/\mathrm{CV}_{R_p, \mathrm{rand}}$. This dispersion measure is the coefficient of variation ($\mathrm{CV} = \sigma/\mu$) of the planet radii within a system normalized by that of an equal number of planets pulled randomly from the population.  

The regressions reveal several weak but statistically significant correlations of planet parameters with $M_{\star}$ and $\mathrm{Fe/H}$; $v \sin i$ and stellar age had no significant effects. Interestingly, the correlations we identified strengthen significantly when considering systems with progressively higher multiplicity; before discussing that phenomenon in more detail, we will first summarize the results for the 128 systems with three or more planets. For $a_{\mathrm{inner}}$, $M_{\star}$ and $\mathrm{Fe/H}$ explain $\sim 22\%$ of the variation ($R^2 = 0.22$), with a positive correlation of $a_{\mathrm{inner}}$ with $M_{\star}$ ($\beta = 0.069 \mathrm{AU}/M_{\odot}$, $p < 10^{-5}$) and a negative correlation with $\mathrm{Fe/H}$ ($\beta = -0.057 \mathrm{AU/dex}$, $p < 10^{-5}$).\footnote{Three outliers with $a_{\mathrm{inner}} > 0.15 \ \mathrm{AU}$ were removed before computation.} The correlations are slightly stronger for $T_{\mathrm{eq, inner}}$; $M_{\star}$ and $\mathrm{Fe/H}$ together describe $\sim 30\%$  of the variation in $T_{\mathrm{eq, inner}}$, and both have positive correlations. 

For the variation in $\mathrm{median}(R_p)$, $\mathrm{Fe/H}$ can alone explain $\sim 21\%$ with a positive correlation ($\beta = 2.2 R_{\oplus}/\mathrm{dex}$, $p < 10^{-7}$, $R^2 = 0.21$). $M_{\star}$ produces no significant impact.\footnote{Four outliers with $\mathrm{median}(R_p) > 5 R_{\oplus}$ were removed before computation.} Likewise, the metallicity is positively correlated with $\mathrm{CV}_{R_p}/\mathrm{CV}_{R_p, \mathrm{rand}}$ ($\beta = 0.7 \ \mathrm{dex}^{-1}$, $p < 0.0005$, $R^2 = 0.1$).

We stress that this dataset exhibits several mild violations of the assumptions of a multiple linear regression, specifically multicollinearity ($\mathrm{Fe/H}$ and $M_{\star}$ are correlated with $R^2 = 0.19$), non-linearity, and heteroscedasticity. To address these, we performed non-parametric tests using Spearman's and Kendall's rank correlations. While there are (expected) differences in the correlation coefficients, all conclusions are consistent with the multiple regression analyses. 

\begin{figure*}
\epsscale{1}
\plotone{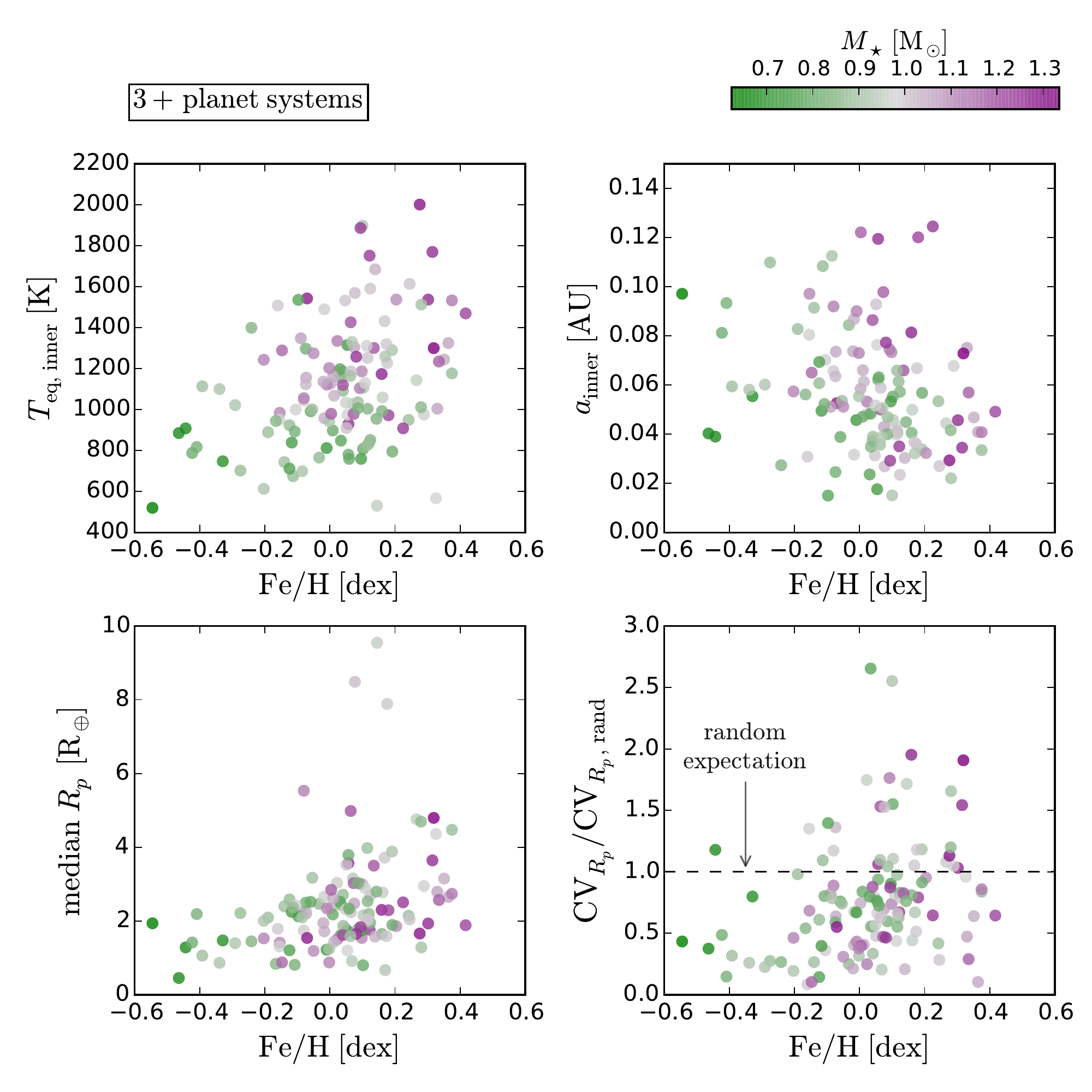}
\caption{Correlations in planetary characteristics of the 3+ \textit{Kepler} multis with $M_{\star}$ and $\mathrm{Fe/H}$. 
The dashed line in the bottom right panel is the expectation of the $\mathrm{CV}_{R_p}/\mathrm{CV}_{R_p, \mathrm{rand}}$ dispersion measure if systems were assembled randomly from the collective population. 
} 
\label{CKS_correlation_summary_3plus}
\end{figure*}

Figure~\ref{CKS_correlation_summary_3plus} displays the correlations graphically. The most interesting features are the following: 
\begin{itemize}
\item[-] System inner edges trend closer to the star for higher metallicities. \textit{This is true despite the positive correlations in $a_{\mathrm{inner}}$ with $M_{\star}$ and in $\mathrm{Fe/H}$ with $M_{\star}$.}
\item[-] The median planet radii tend to increase with metallicity. 
\item[-] The within-system radius dispersion is generally smaller than random expectation (as already discussed in Section~\ref{clustering}). When size dispersion is large, it is typically associated with high metallicities, $\mathrm{Fe/H} \gtrsim 0$. 
\end{itemize}

As mentioned previously, the correlations strengthen significantly with planet multiplicity. For example, the $R^2$ of the correlation between $M_{\star}$ and $\mathrm{Fe/H}$ (as independent variables) and $a_{\mathrm{inner}}$ (as the dependent variable) is 0.1 for systems with 2+ planets, 0.22 for $3+$, and 0.39 for $4+$. By sampling subsets of the $2+$ systems, we confirmed that this observation is not simply a consequence of the decrease in sample size with multiplicity.

\section{Discussion}

We expanded on the findings of \cite{2017arXiv170606204W} to investigate the within-system dispersion in the planet masses and radii of $\textit{Kepler}$ multi-transiting systems. The conclusions are striking: The masses and radii in a given system are significantly more uniform as compared to random expectation (to $\sim 8.3\sigma$). Moreover, both masses and radii tend to be ordered when the planets are ranked by orbital period. In light of these new observations, the large astrophysical scatter of the mass-radius relation has suddenly been endowed with much more structure.

It is important to point out the selection biases that may be affecting our results. We considered a sample of 37 \textit{Kepler} multis with TTV masses measured by \cite{2017AJ....154....5H}. The covariance between planet masses in the TTV posteriors (which we mentioned following equation~\ref{dispersion metric}) could work to slightly boost the observed phenomenon of intra-system mass uniformity. However, the correlations are not always positive, particularly for systems with more than two planets. Furthermore, our results are strictly only relevant for near-resonant systems with sizeable TTVs. 
It is conceivable that these systems have planet masses that are significantly more uniform than the broader population of \textit{Kepler} close-in, coplanar systems. However, this seems unlikely given that the tendency for systems to exhibit radius uniformity exists beyond this TTV sample; it would be unusual for the planets in a system to coordinate their radii but not their masses. 

The correlations that we have found allow us to sketch a plausible framework of explanation, while keeping in mind that planet formation is likely a highly nonlinear process, and that \textit{any} such argument cannot be construed as strictly deterministic.  

\cite{2015ApJ...811...41L} outlined an analytic framework that links the planetary gas-to-core mass ratio (GCR) of a planet to protoplanetary nebular variables. For disk lifetime, $\tau$, and gas metallicity, $Z$, their theory predicts GCR $\propto \tau^{0.4}M_{\rm core}^{1.7}Z^{-0.4}(1-Z)^{-3.4}$. In other words, enhanced cooling associated with a high metallicity envelope can permit higher GCRs and hence larger planets. This insight, together with the observational evidence that disk lifetimes are shorter in low metallicity environments \citep{2010ApJ...723L.113Y}, is consistent with the positive correlation between the median system radii and  stellar metallicities shown in Figure~\ref{CKS_correlation_summary_3plus}. 

For disks with surface density power-laws concordant with the Minimum Mass Extrasolar Nebula \citep{2013MNRAS.431.3444C}, solid body accretion simulations, \citep[e.g.][]{2013ApJ...775...53H}, show the rapid development of multiple solid cores with similar masses. A set of similar cores will accrete gas at comparable rates, thereby finishing the formation process with roughly equal masses and radii. The enhanced cooling at high metallicity, however, makes it more likely for some planets to approach runaway accretion, resulting in greater intra-system size dispersion. Furthermore, the central role of metallicity as a coolant permits planet formation at high equilibrium temperature, which may contribute to the correlations that we have noted.

\section{Acknowledgements}
We thank the anonymous referee for insightful comments and suggestions. S.M. is supported by the NSF Graduate Research Fellowship Program under Grant  DGE-1122492. S.W. acknowledges the Heising-Simons Foundation for their support. G.L. acknowledges NASA Astrobiology Institute support under Agreement \#NNH13ZDA017C.


\end{document}